# Reduced work function of graphene by metal adatoms

Merid Legesse[1], Fedwa El Mellouhi[1], El Tayeb Bentria[1], Mohamed E. Madjet[1], Timothy S Fisher[2], Sabre Kais[1, 3,4], Fahhad H Alharbi[1,4,*]

[1]Qatar Environment and Energy Research Institute, Hamad Bin Khalifa University, Doha, Qatar.

[2]School of Mechanical Engineering and Birck Nanotechnology Center, Purdue University, West Lafayette, Indiana 47907, USA.

[3]Department of Chemistry and Physics, Purdue University, West Lafayette, Indiana 46323, USA.

[4]College of Science and Engineering, Hamad Bin Khalifa University, Doha, Qatar.

[*]Email: falharbi@qf.org.qa

## Abstract

In this paper, the work function of graphene doped by different metal adatoms and at different concentrations is investigated. Density functional theory is used to maximize the reduction of the work function. In general, the work function drops significantly before reaching saturation. For example in the case of Cs doping, the work function saturates at 2.05 eV with a modest 8 % doping. The adsorption of different concentrations on metal adatoms on graphene is also studied. Our calculations show that the adatoms prefer to relax at hollow sites. The transfer of electron from metallic dopants to the graphene for all the studied systems shifts the Fermi energy levels above the Dirac-point and the doped graphenes become metallic. The value of Fermi energy shifts depends on the type of metallic dopants and its concentrations. A detail analysis of the electronic structure in terms of band structure and density of states, absorption energy, and charge transfer for each adatom-graphene system is presented.

# 1. Introduction

Graphene is a flat monolayer of carbon atoms tightly packed into a two-dimensional (2-D) honeycomb lattice and constitutes the basic building block for graphitic nanostructures of varying dimensionalities. It can be wrapped up into zero-dimensional (0-D) fullerenes, rolled into one-dimensional (1-D) nanotubes or stacked into three-dimensional (3-D) graphite [1-3]. Its exceptional electric and transport properties have attracted wide attention since its discovery. Also, it exhibits extraordinary electronic structure where the $\pi$ and $\pi^*$ bands merge into a single point at the Fermi energy (Dirac point) and display a linear $E$ vs. $k$ dispersion. Such a unique feature implies that the electrons at the Fermi level ($E_F$) in graphene behave like massless relativistic particles and leads to some interesting properties such as the anomalous quantum Hall effect [4-6], high electron mobility at room temperature [2,6], and exceptional thermal and mechanical [7,8] properties. Because of such extraordinary characteristics, it is expected that graphene will have great potentials in electronic devices and many other applications [1,9-18].

However, there are still many obstacles for graphene to overcome before being adopted as a device material. Recently, graphene doping have drawn much interest because it is crucial to fabricate integrated devices [19]. Thus, metal adatoms on graphene have been a topic of great interest since they can locally dope or modify its band structure [20-23]. The interaction of electrons in graphene with surface metal adsorbates is a desirable feature providing high electronic mobility, doping, and some other particular applications pertinent to sensing applications [9-12]. Furthermore, the adsorption of metal nanoparticles is known to change the structural and electronic properties of graphene [24,25] and might be employed as an electrical contact as well, hence the importance of fully understanding their impact. [26]

Alkali metals (highly reactive chemical species) have a fairly simple electronic configuration, and their adsorption on a graphite surface have been investigated for the past 24 years [27]. Metal-graphene, metal-graphite and metal-nanotube structures are of particular interest for catalysis, fuel cell technology, hydrogen storage, metal-ion batteries and solar cells [9-18]. Because of the diversity of properties of the metallic elements, the variety of structures formed, and the availability of experimental techniques at the nanoscale, the adsorption of metals on the surface of this material is a promising approach to controllably modify graphene work function as well as electronic properties.

In graphene-based electronic devices, the graphene work function is one of the most important properties that should be taken into considerations. In graphene electronics, the contact resistance between graphene and metal limits the performance of devices due to its effects on carrier mobility [28]. Contact resistance suppresses the ON-current, which is determinant to high-frequency transistor performance; however, in optoelectronics this effect can enhance the photocurrent [29], as lower work function can dramatically enhance the emitting current. The work function of graphene electrodes is also critical to maximize energy conversion efficiency in solar cell. Besides good conductivity and transparency of the electrode, the performance and current density for semiconducting electronic devices such as light-emitting diodes and field-effect transistors depends strongly on the carrier injection efficiency through the contact between electrodes and semiconducting material layers. Furthermore, graphene is considered a prospective conductive material with the desirable properties including engineered work function for efficient carrier injection. Different approaches have been investigated to tune the work function of graphene including reaction with organic and inorganic molecules [30-33], chemical modification of the surface [34], metal doping [23,35,36], substrate orientation [37], self-assembled monolayer formation [38], and external electric field [39]. Intercalation of different species such as H [40], F [41], Li [42], Au [43] and Fe (III) Cl [44] has also been identified as a potential method to tune the work function of few-layer graphene [44]. However, controlling the work function of graphene precisely and on demand has yet to be demonstrated. In this paper using density functional theory (DFT), we targeted a systematic reduction of work function by varying the concentration of different metal adatoms (Adatoms: Li, Na, K, Rb, Cs, Be, Ca, Mg, Sr and Ti). By heavily doping Cs, Rb, K, Na, and Li on the graphene sheet we compute the lowest saturated values for the work function of graphene which importantly is in good agreement with the experimentally reported value in the case of K-doped graphene [45].

## 2. Computational details

In this work, Density Functional Theory (DFT) calculations are conducted to study the electronic structure properties and work function of metal-adatom doped graphene (Li, Na, K, Rb, Cs, B, Ca, Mg, Sr and Ti) at different concentrations. All electronic structures and related properties such as the work function reported herein were performed using the projector augmented wave (PAW) [46, 47] method as implemented in VASP [48]. The exchange-correlation interaction was treated in the generalized gradient approximation (GGA) in the parameterization of Perdew, Burke, and Ernzerhof (PBE) [49] where the kinetic-energy cutoff was set to 400 eV. A 8x8x1 Monkhorst-Pack mesh [50] for the Brillouin zone sampling and a Gaussian broadening of 0.2 eV were employed. Band structure was calculated along the Γ–M–K–Γ k-path. A semi-empirical van der Waals correction term vdW-DF [51-54] was included which is known to have a significant impact on the adsorption energies on graphene surfaces.

A supercell technique was adopted for the calculations with varying surface areas (variable number of carbon atoms) and a vacuum region equivalent to 12 Å. We validated the vacuum region based on the average electrostatic potential plot. The flat vacuum region in the average electrostatic potential plot is one of the indications for the vanished interaction between the supercells along the vacuum direction. In other word, the vacuum region is wide enough. We started the calculation from small vacuum region (8 Å) and then increase it until we obtained the flat vacuum region in the average electrostatic potential graph. The graph starts flattening when the separation is around 12 Å. The calculations were repeated with 14 and 16 Å vacuum region leading to the same results. To reduce the computational cost, we keep the optimal 12 Å vacuum region for the entire calculations. The equilibrium atomic positions were determined by relaxing the entire system with fixed lattice vectors. The above computational settings yielded electronic properties in excellent agreement with those reported experimentally.

## 2.1 Work function and adsorption of graphene with adatoms

The work function ($\Phi$) of any material can be defined as the energy required to remove an electron from the highest filled level in the Fermi distribution of a solid to vacuum (i.e., stationary at a point in a field-free zone just outside the solid) at absolute zero temperature. The $\Phi$ of doped graphene is defined as the energy difference between the vacuum and $E_F$,

$$\Phi = E_{vac} - E_F \tag{1}$$

where $E_{vac}$ is the vacuum level and $E_F$ is Fermi level. In the numerical calculations, $E_F$ is determined by integrating the density of states from the lowest energy level to an energy level that gives the total number of electrons in the unit cell. $E_{vac}$ is obtained using the planar-average electrostatic potential energy along the $z$ direction (vacuum direction) [55].

For illustration the average electrostatic potential of 7x7 pure and K-doped graphene are plotted in Fig. 1. As apparent in Panel (a), the calculated Φ of pure graphene using equation 1 is 4.38 eV and is very close to the experimentally measured Φ of graphene which ranges between 4.45-4.5 eV [30, 56, 57]. For Panel (b), the calculated Φ in K-doped graphene is 3.22 eV. Using the same approach, we have calculated the Φ of graphene doped by various metal adatoms (see Section 3).

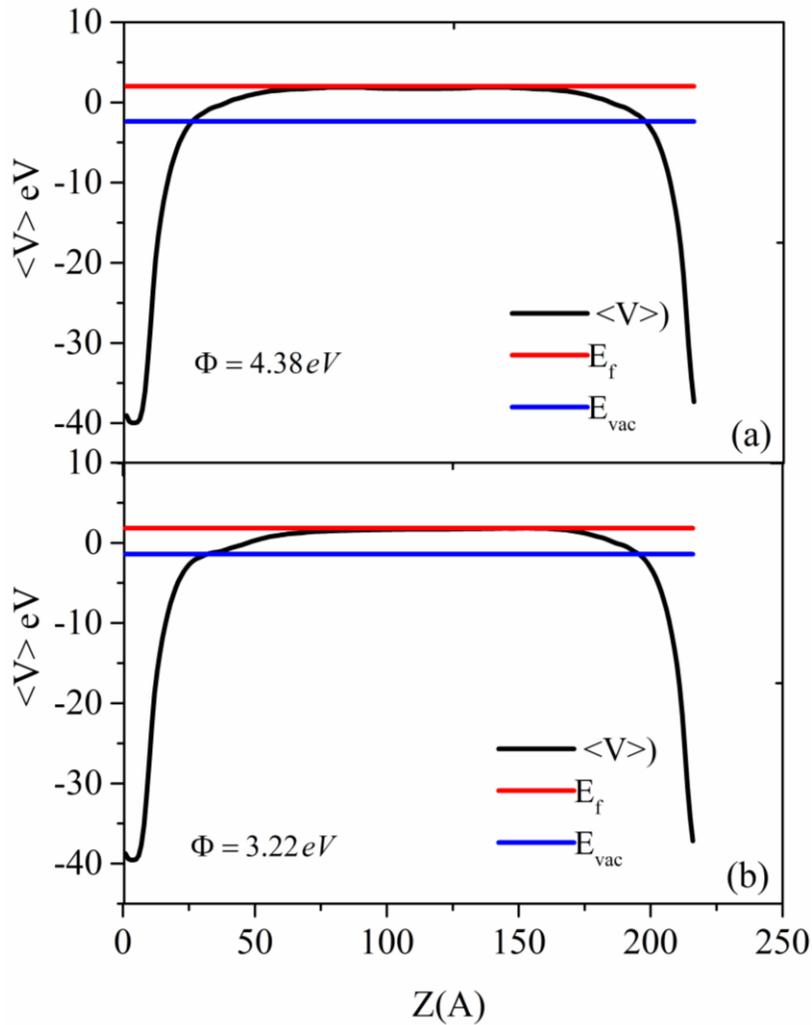

**Figure-1** The electrostatic potential $\langle V \rangle$ averaged over the $xy$ plane as function of the vacuum direction $z$ (Å) for single layer (a) pristine graphene (b) K-doped graphene.

As for the adsorption of the adatom with the graphene, it is assessed by the calculated binding energy, defined as :

$$E_b = \frac{E_G + nE_M - E_{G-nM}}{n} \qquad (2)$$

where $E_{G-nM}$ is the total energy of the graphene sheet with metal, $E_G$ is the total energy of the pristine graphene sheet, $E_M$ is the total energy of the free metal adatom, and $n$ corresponds to the number of metal adatoms.

## 2.2 Structural analysis of graphene with metal adatoms

The optimization of the metal-graphene system is performed as follows: First, the geometries of graphene sheets are optimized, then doped with metals adatoms occupying the hollow (H) sites (at the center of a hexagon), on top (T), and finally bridge (B) between two carbon atoms and above certain height. The distances between the metal adatoms and the graphene sheet were set to values slightly smaller than the sum of the adsorbate and carbon atoms covalent radii. For each adsorption site, the adatom is relaxed along the $z$ direction (vacuum direction) while the C atoms' internal coordinates are relaxed in all directions until forces on them become less than 0.01 eV/Å. In all cases, the adatoms prefer to relax at the hollow site which is consistent with experimental and other theoretical studies [23, 58-60]. For further calculations, we have only focused the hollow site structural configurations of metal doped graphene. In Fig. 2, the relaxed structures of metal adatoms on 7x7 graphene sheets (98 C atoms) are shown with increasing concentration of metallic adatoms.

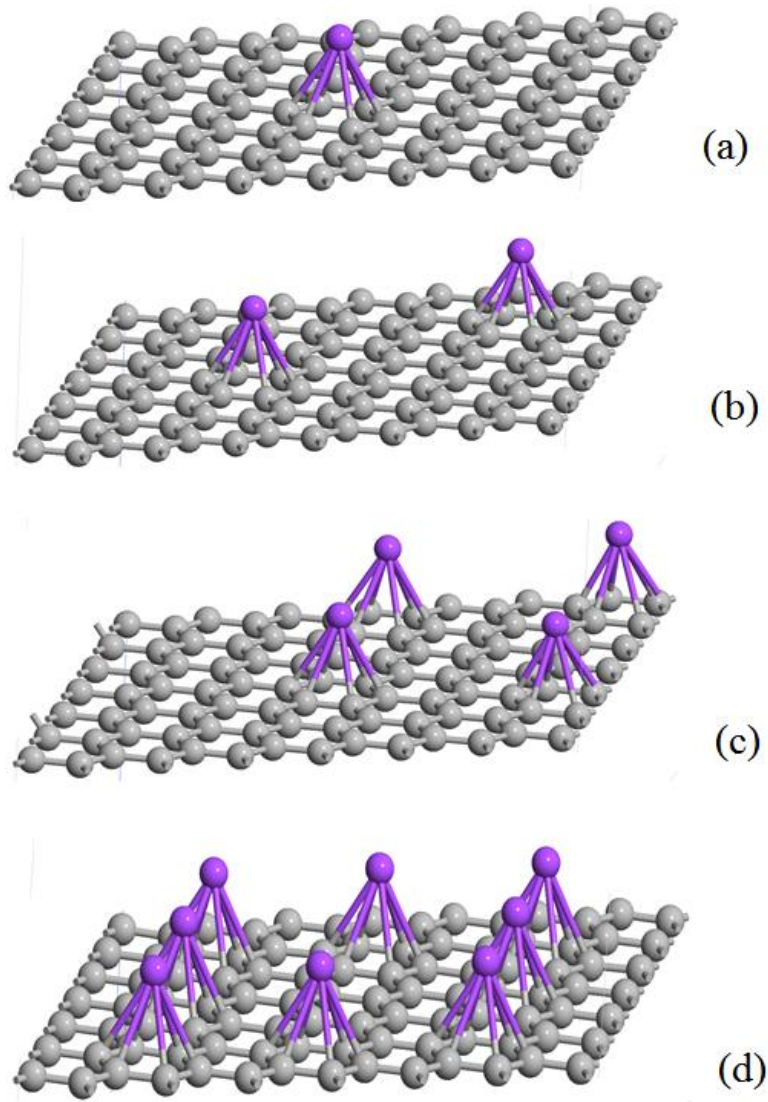

**Figure-2** The relaxed structures of a 7x7 graphene sheet with metal adatoms. (a) one adatom (b) two adatoms (c) four adatoms, and (d) eight adatoms.

## 3. Results and discussion

First, the adsorption characteristics of adatoms are investigated, starting with alkali metals known to be highly reactive with increasing chemical activity while moving from Li to Cs. The resulted structures from full geometry optimization show indeed that all the studied alkali atoms

(Li, Na, K, Rb and Cs) favour bonding on the hollow sites of the graphene sheet. This results in small distortion or strain within the graphene sheet. The characteristic bonding geometries of alkali atoms are summarized in Table-I. Here, the height of adatoms is calculated as the difference between the average coordinates of neighboring C atoms and the adsorbate. The equilibrium distance between the adatom and graphene sheet monotonically increases with increasing atomic size. These results are consistent with those in the literature [58, 61-66].

**Table-I** The binding energy of adatoms ($E_b$) and their equilibrium height above graphene sheet ($d_{M-G}$)

| Graphene Adatom | $d_{M-G}$ (Å) | Binding energy $E_b$ (eV) |
|---|---|---|
| Li | 1.91 | 1.68 |
| Na | 2.41 | 0.05 |
| K  | 2.82 | 0.70 |
| Rb | 2.98 | 0.88 |
| Cs | 3.13 | 1.20 |
| Be | -    | -3.95 |
| Mg | -    | -1.65 |
| Ca | 2.48 | 1.13 |
| Sr | 2.71 | 0.81 |
| Al | 2.28 | 2.41 |
| Ti | 2.47 | 1.46 |

In the case of alkali earth metals, the distance between the adatom and graphene sheet monotonically increases with atomic size except for Be and Mg. In fact, the negative binding energies of both species indicate that they are physisorbed and likely to desorb from the graphene sheet. Therefore, the interaction distance between Be and Mg with graphene sheet isn't defined. The physisorption character of Mg could be well understood by its ionization potential of 7.64 eV, which is higher than that of Ca (6.11 eV) and Sr (5.69 eV) [55]. The electrons in s orbitals of Mg cannot be easily transferred to the graphene. We therefore conclude that Mg does not interact with graphene sheet as also suggested before by other theoretical works [66-67]. This is shortly illustrated by electron charge density distribution (Fig. 3d). Like Mg, the interaction between Be and graphene sheet is very weak and Be also exhibits a physisorption character. The ionization potential of Be is 9.32 eV [55], which is higher than all alkali earth metals.

In order to visualize the resulted charge density distribution of the doped graphene, the calculated charge density distribution is plotted in Fig. 3. Panel (a) clearly shows the strong interaction between Cs and graphene sheet. A large portion of the electron density cloud of Cs is transferred and distributed throughout the graphene surface. This indicates that the interaction between Cs and graphene is strongly ionic. In contrast and as shown in Fig. 3 (d), the interaction between Mg and graphene is almost negligible as it is physisorption (as discussed above). The charge density distribution of Al and Ti is also plotted in Fig. 3 (b and c). For Ti case, the interaction is weaker and the Ti-donated electron charge density is highly localized around the interacting carbon atoms. Essentially, the resulted charge density distributions are consistent with quantitative analysis which presented in Table I.

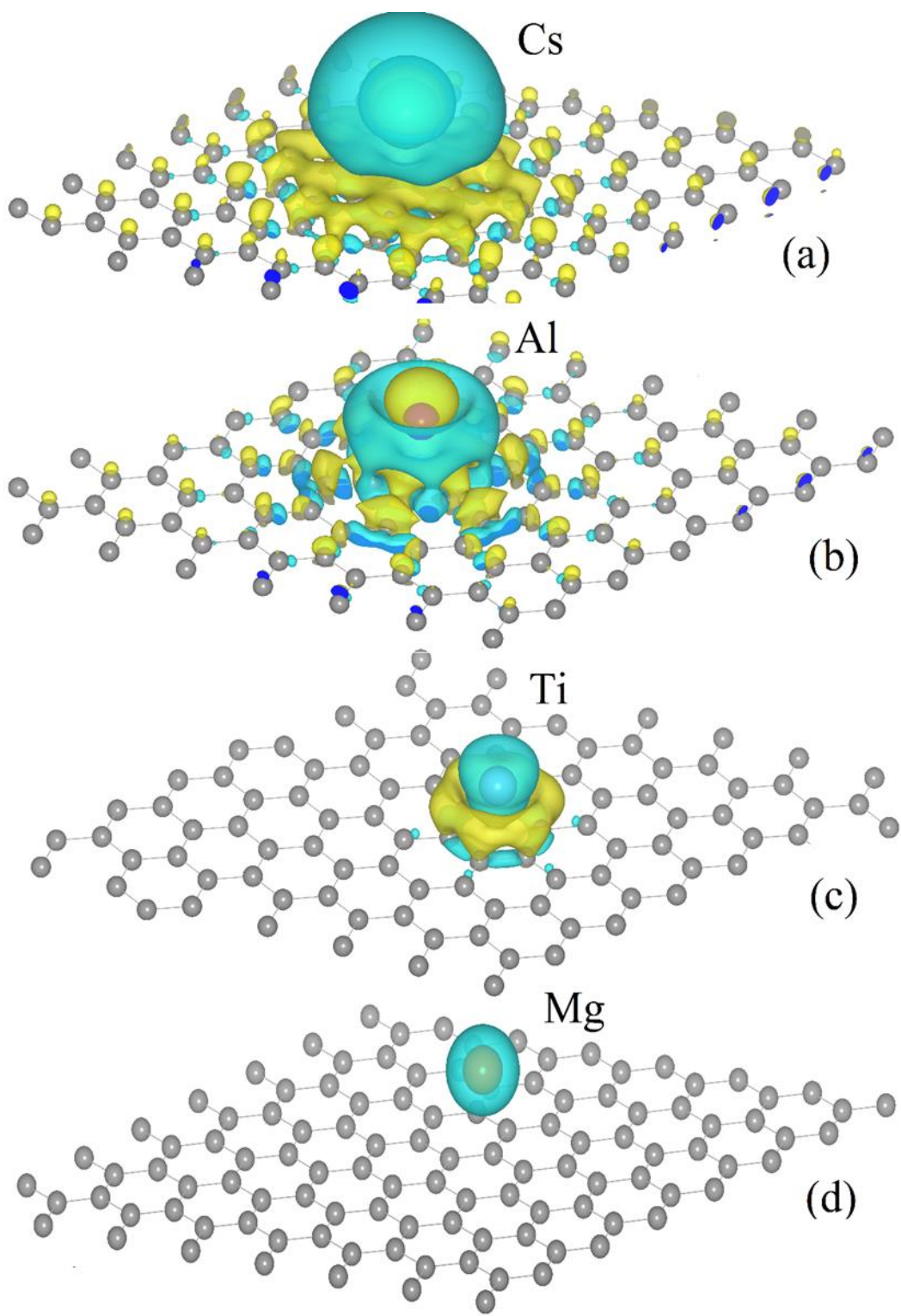

**Figure-3**  Electron charge density distribution of graphene-doped (a) Cs, (b) Al, (c) Ti and (d) Mg.

For band structure calculations, the conventional representation of the energy dispersion relations along the lines between high symmetry points Γ, M, K and Γ of the first Brillouin zone in reciprocal space is used. In this case, we have taken 90 k-points along the specific direction of the irreducible Brillouin zone to obtain fine band structure. Graphene have many interesting properties because of two dimensionality. Electrons in single-layer graphene exhibit a characteristic linear dispersion relation between energy and momentum near the K-point. The band structure of pristine graphene is shown in Fig. 4 (a). In this figure you can see that the $E_F$ crosses on the Dirac point (the point, where the π bond touches the anti-bond π* of carbon atoms) which is typically characteristic of graphene. In Fig. 4-7, the electronic band structures of Rb, Ca, Al and Ti adsorbed on graphene are shown as an example, and their properties fairly represent the overall adatom-graphene system. In general, the electronic band structure calculations indicate that the adsorption of Rb, Ca, Al and Ti atoms on graphene results in a shift of the $E_F$ above the Dirac point.

The calculated band structure for various concentrations of Rb adsorption on graphene sheets are presented in Fig. 4. As a result of Rb adsorption, the charge donated from 5s orbital of Rb into 2p orbital of carbons. The unoccupied 2p (antibonding π*) states become occupied and then the C-C bond becomes distorted. These distortions of C-C bond changes the zero-gap semiconductor behaviour of graphene into metallic. By controlling the concentration of metal adatoms, we can tune the position of $E_F$ in the linear region of bands crossing at the K-point of the Brillouin zone.

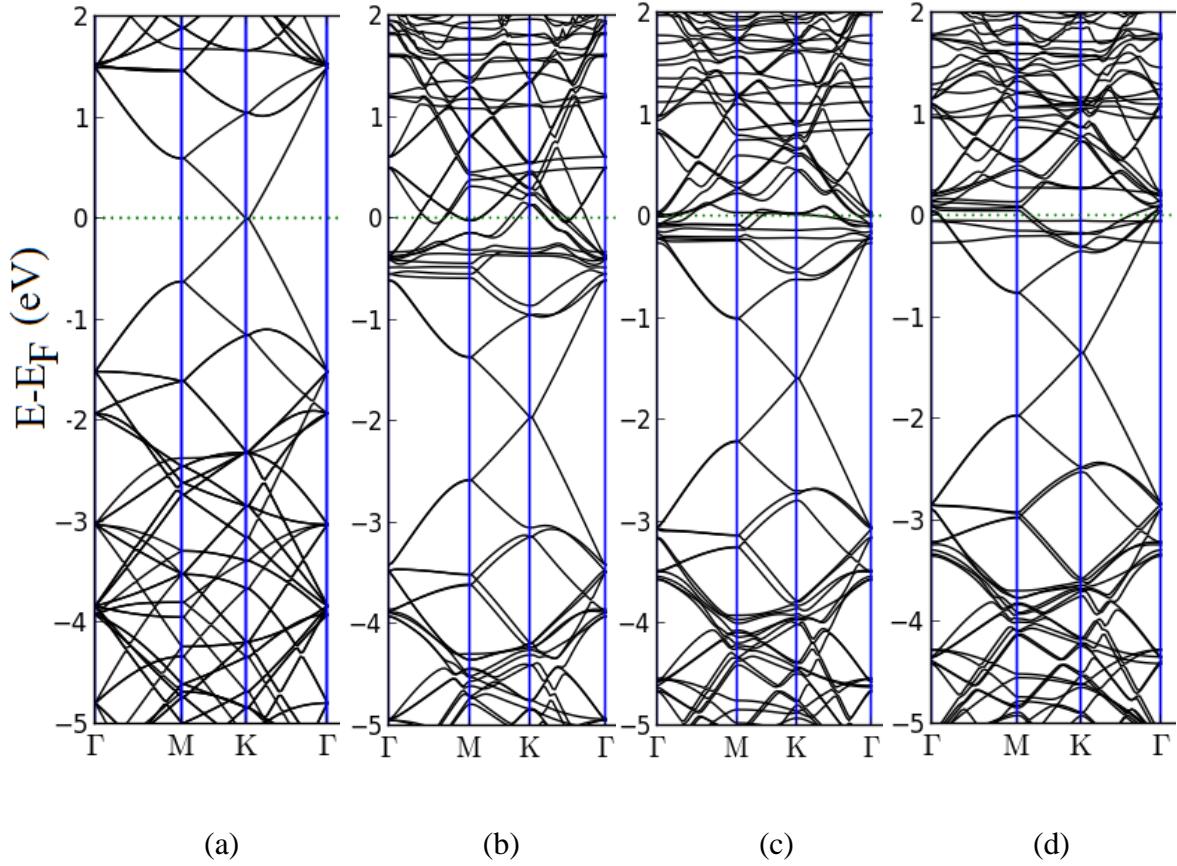

**Figure-4** Band structures of (a) pristine graphene and graphene doped by (b) one-Rb atom, (c) two-Rb atoms, and (d) four-Rb atoms. The reference energy is to $E_F$.

$E_F$ for different concentrations of Rb adatom lies in different levels above the Dirac point. These variations of $E_F$ are due to the fact that the transfer of electrons from adatoms with different concentrations to graphene sheet is different. When we change the concentration Rb from pristine graphene to 4 adatom on graphene sheet, $E_F$ also changes from -2.36 to -0.30 eV. This results in the reduction of $\Phi$.

By moving to the second group, the band structure of the Ca-doped graphene at different doping concentrations are shown in Fig. 5 (a-c). It is clear that all the three configurations exhibit metallic character. The $2p$ states of C crosses the $E_F$ due to the electron transfer from adatoms to graphene. Therefore, the interactions between the adatoms and graphene are ionic.

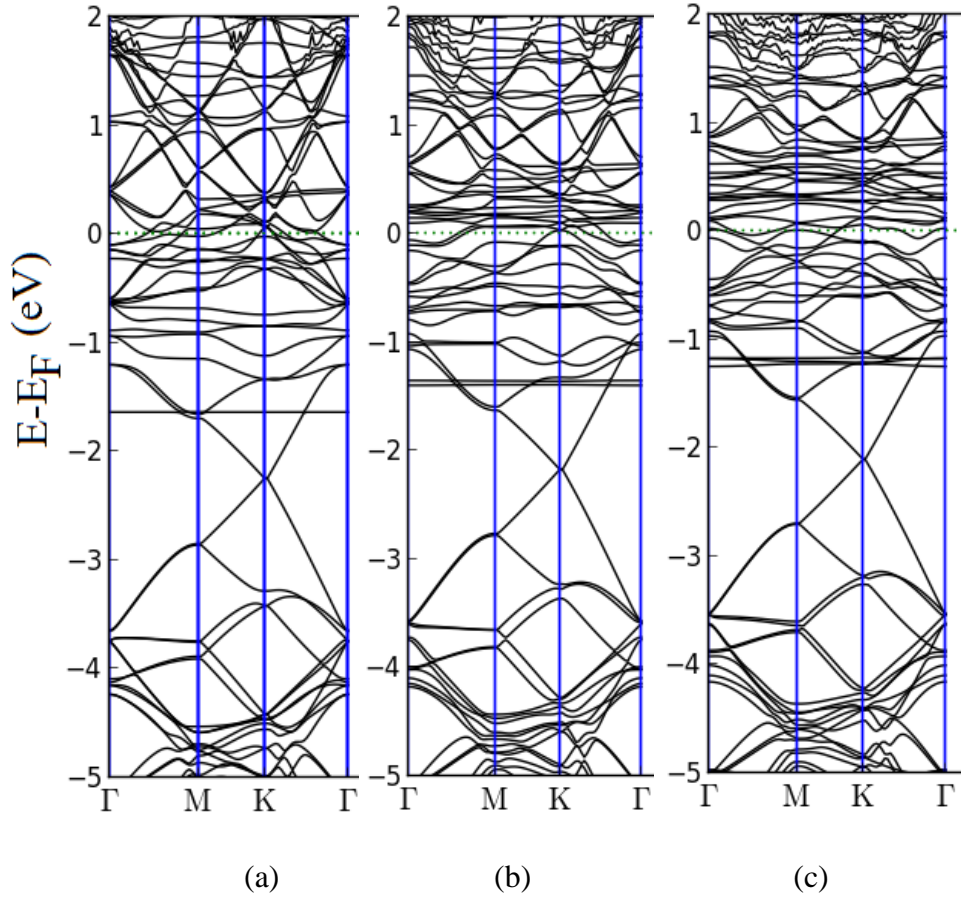

**Figure-5** Band structures of graphene doped by (a) one Ca atom, (b) two Ca atoms and (c) four Ca atoms. The reference energy is to $E_F$.

To complete the analysis by considering transition metals (Al and Ti are considered), the band structure of Al and Ti-doped graphene resulting from various concentrations are shown in Fig. 6 and Fig. 7 respectively. The trend of electronic properties Al- doped graphene is similar to Ti-doped graphene. As can be observed from the band structure in Fig. 7 and the corresponding PDOS (Fig. 8) of Ti-doped graphene, the Ti 3d states hybridized with the C 2p states below and above the $E_F$. This interaction is different from alkali metal doped graphene where simply the extra electron from metallic dopants transfers to the 2p states of carbon. Unlike in Ti case where the interaction depends on hybridization between Ti 3d and 2p states of C. Due to this reason, the Ti-donated electron is shared between Ti and the interaction carbon atom and hence it is highly localized (Fig. 3 (c)). Thus, $E_F$ shift from the Dirac-point is small compared to alkali-metal doped graphene.

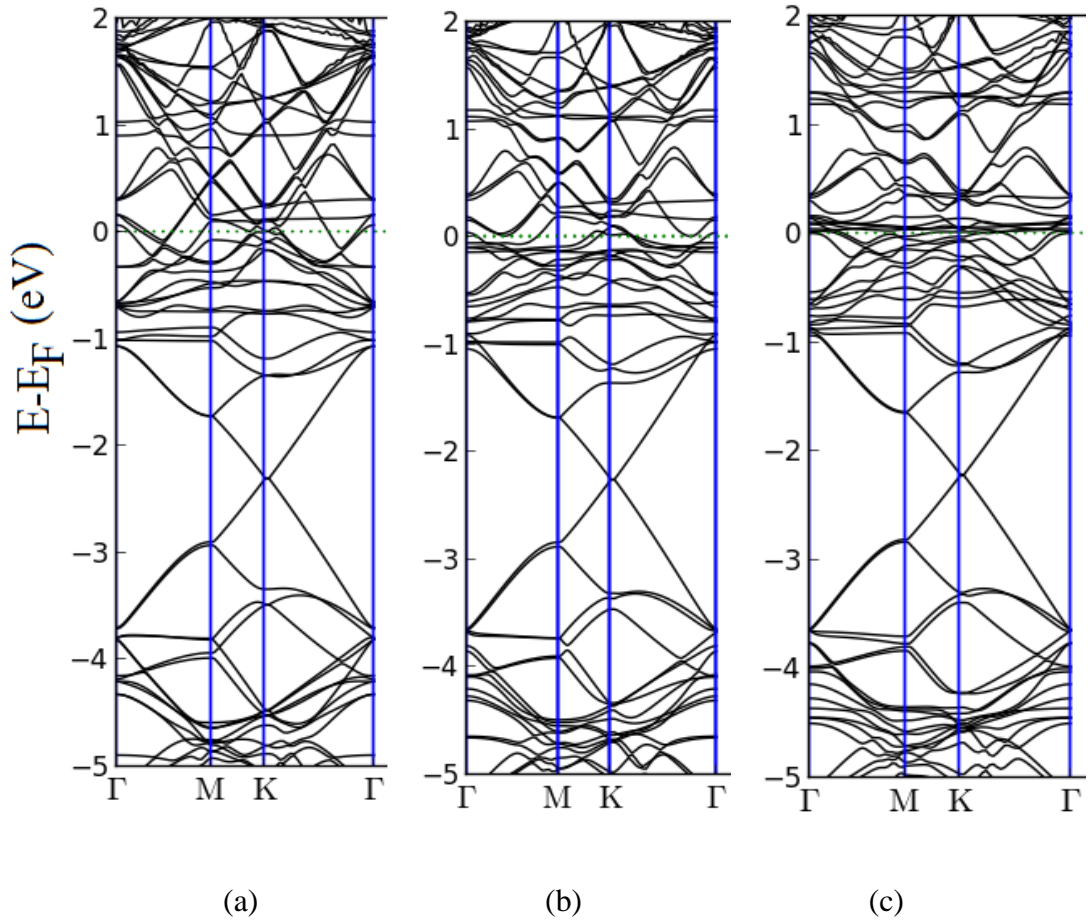

(a)          (b)          (c)

**Figure-6**    Band structures of graphene doped by (a) one Al atom, (b) two Al atoms and (c) four Al atoms. The reference energy is to $E_F$.

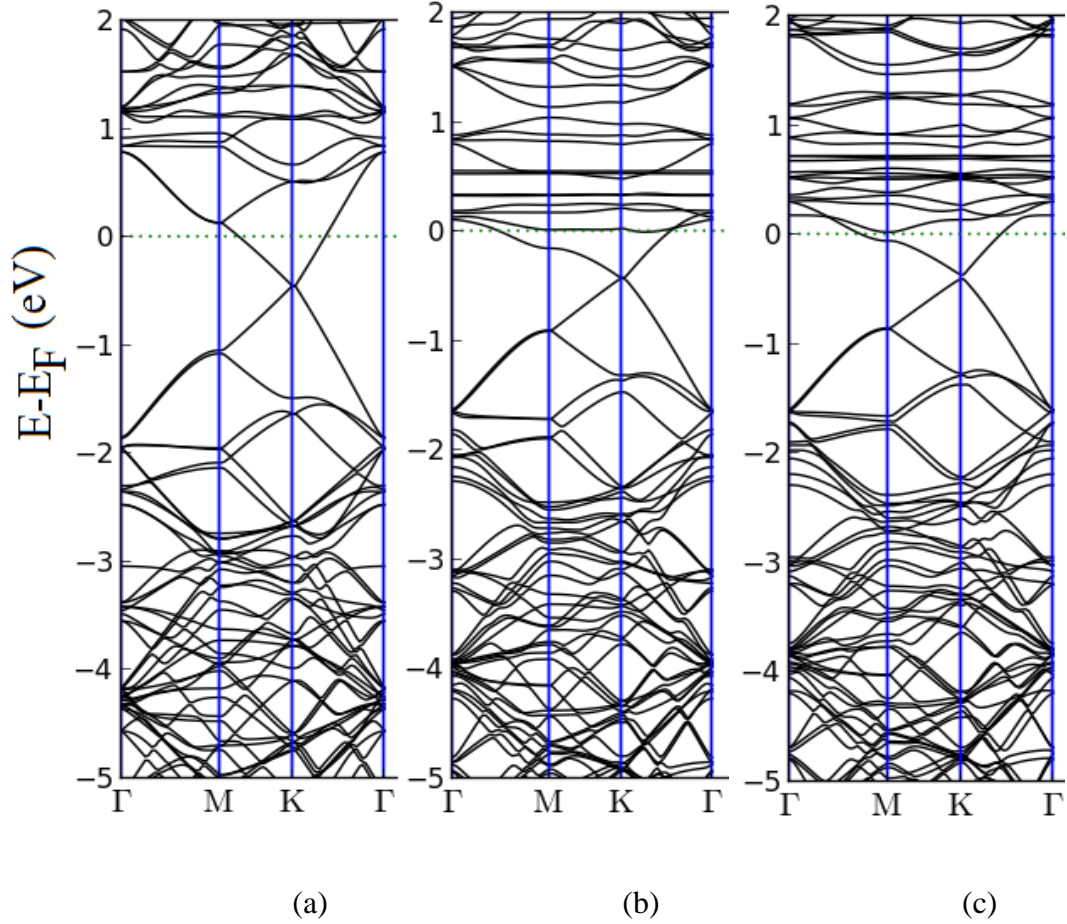

|       (a)       |       (b)       |       (c)       |

**Figure-7**   Band structures of graphene doped by (a) one Ti atom, (b) two Ti atoms and (c) four Ti atoms. The reference energy is to $E_F$.

In the next step, the projected density of states (PDOS) for the equilibrium geometry of adatom-graphene was computed. The PDOS of single adatom Rb, Ca, Al and Ti-doped graphene are plotted and presented in Fig. 8. In Fig. 8 (a), the Rb 5*s* state lies above $E_F$ and it is unoccupied, this indicates that a very large charge transfer takes place between the Rb and graphene. The result suggests that Rb-graphene bonding is of ionic nature and 0.95 eV electron transfer from the *5s* state of Rb to 2p sates of graphene.

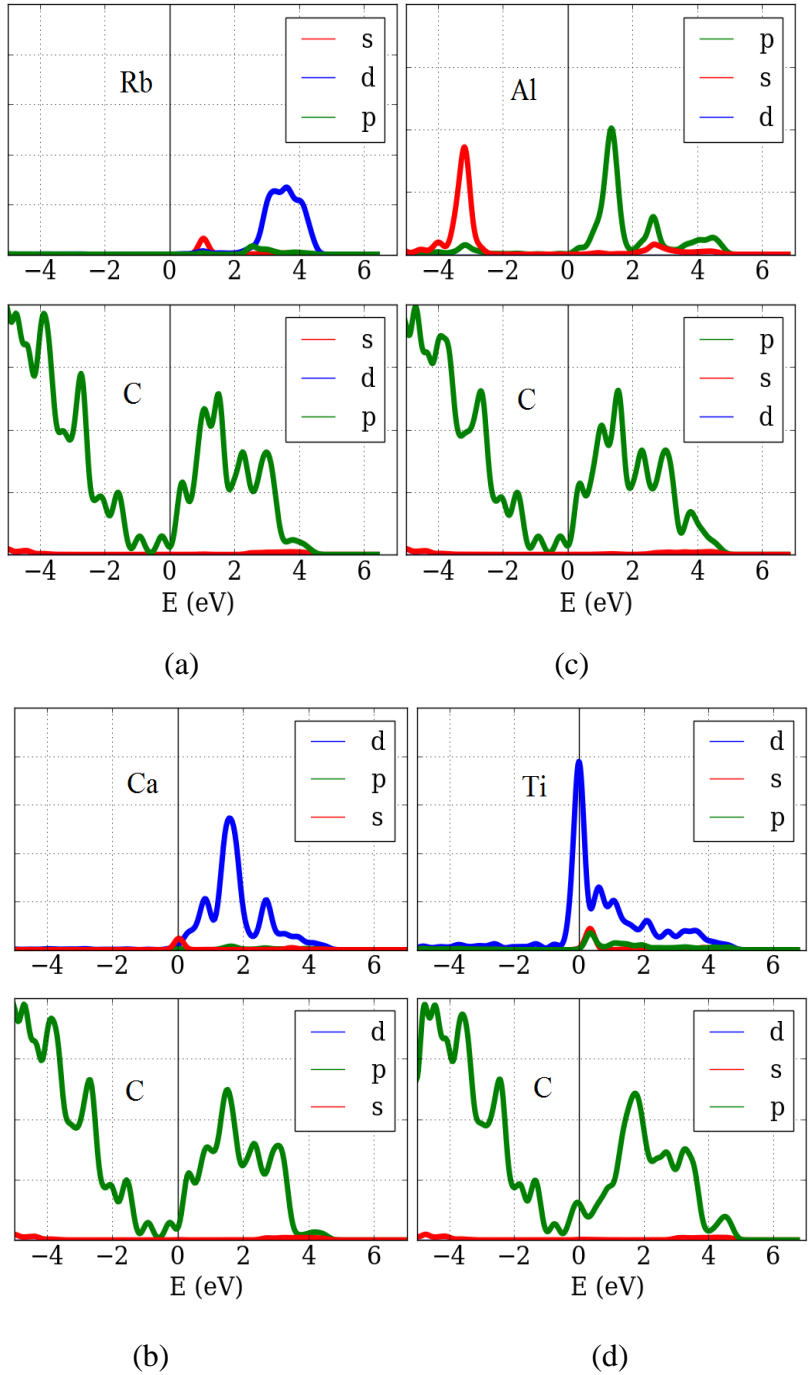

**Figure-8** Projected density of states (PDOS) of (a) Rb-doped graphene, (b) Al-doped graphene, (c) Ca-doped graphene and (d) Ti-doped graphene. The reference energy is $E_F$. The density of states presented in bottom panels is projected to C p states and upper panels adatom s, p and d states (shown as red, green and blue solid line respectively).

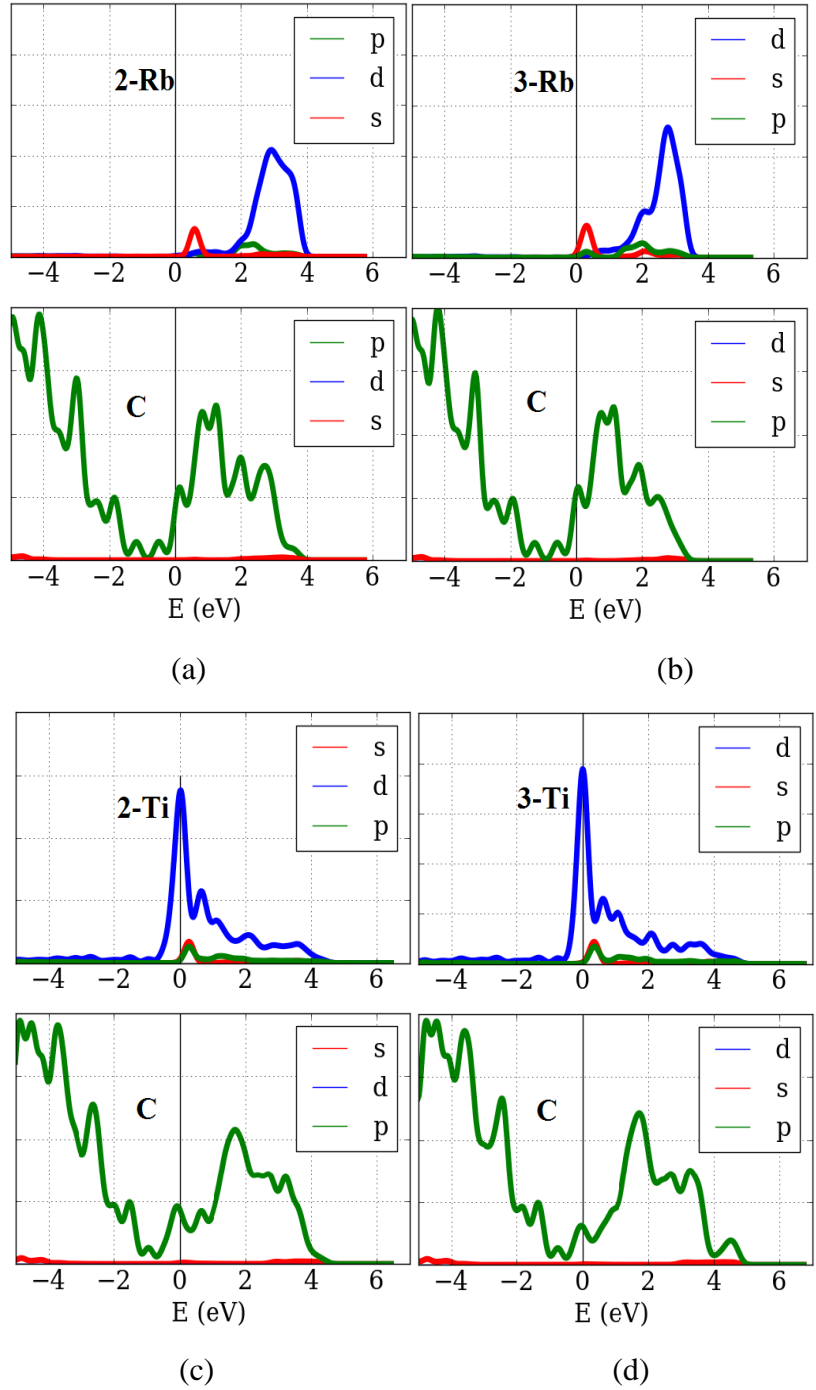

**Figure-9** PDOS of (a) two Rb-doped graphene, (b) three Rb-doped graphene, (c) two Ti-doped graphene, and (d) three Ti-doped graphene. The reference energy is $E_F$. The density of states presented in bottom panels projected to C p states and upper panels adatom s, p and d states (shown as red, green and blue solid line respectively).

For Al-doped graphene PDOS, similar to Rb-doped graphene PDOS, the p states of Al lie above $E_F$. However, Al-graphene bonding is weak compared to Rb-graphene bonding because electron transfer from the 3p state of Al to 2p graphene is 0.3 eV. In Fig 8 (c, d) the PDOS states of Ca and Ti-doped graphene is shown. In both cases the $E_F$ lies on p and d states of Ca and Ti adatoms respectively. As a result unlike Rb and Al-doped graphene, higher concentration of Ti and Ca-doped graphene result in a smaller shift of $E_F$ toward the conduction band.

Different concentration of Rb and Ti-doped graphene PDOS is shown in Fig. 9 (a-d). For Rb (see Fig. 9 (a, b)), it clearly shows that the shifting of $E_F$ toward the conduction band upon addition of more concentration Rb adatom. Therefore, the more electron from Rb adatom transfer to graphene sheet. The consequence of shifting the $E_F$ to conduction band is the reduction of $\Phi$.

The full results of the calculations are summarized in Table II and Fig. 10 for alkali metals. All the studied metal-doped graphene examples show a reduction of $\Phi$. However, the reduction varies for each metal adatoms and its concentration. Specially, alkali metals doped graphene reduce the $\Phi$ significantly compared to other studied metal adatom as shown in Table II. The charge transfer is quantified by the difference of the partial charges ($Q_{M-G}$) between the graphene and the metallic adatom. The partial charges are calculated by the Density Derived Electrostatic and Chemical charges method (DDEC) [68, 69]. The resulted $Q_{M-G}$ are shown in Table II. For alkali metals, $Q_{M-G}$ increases as we go down in the group for 0.90 (Li) to 0.98 (Cs). This is because within the group, the ionization energies decrease due to the increasing atomic size in their respective periods. In large atoms, the valence electrons are loosely held by the nucleus and are easily lost. These high values suggest that electrons transfer easily from alkali metals to graphene sheet. For alkali earth metals, the same trend is observed but with smaller charge transfer. However, Mg and Be have the lowest $Q_{M-G}$ of 0.09 and 0.11 respectively. This is because the ionization potential of Mg and Be are higher than all alkali earth metals. For Al and Ti the $Q_{M-G}$ are 0.3 eV and 1.06 eV. Al has low $Q_{M-G}$ and the interaction weakly ionic. But Ti has high $Q_{M-G}$. However, the interaction with graphene is weak this is due to the fact that the adsorption Ti-doped graphene is very low.

**Table-II** Shows the work function ($\Phi$) of different concentrations of metal-doped graphene and charge transfer from adatoms to graphene ($Q_{M-G}$)

| 8 X 8 Dopant | $\Phi$ | | | | | $Q_{M-G}$ |
| --- | --- | --- | --- | --- | --- | --- |
| | 1:98 (eV) | 2:98 (eV) | 3:98 (eV) | 4:98 (eV) | 8:98 (eV) | |
| Li | 3.53 | 3.07 | 2.82 | 2.63 | 2.57 | 0.90 |
| Na | 3.34 | 2.75 | 2.54 | 2.46 | 2.45 | 0.92 |
| K  | 3.22 | 2.49 | 2.25 | 2.23 | 2.23 | 0.93 |
| Rb | 3.15 | 2.34 | 2.21 | 2.20 | 2.18 | 0.95 |
| Cs | 3.04 | 2.26 | 2.06 | 2.06 | 2.05 | 0.98 |
| Ca | 3.46 | 3.24 | 2.92 | 2.91 | 2.91 | 0.68 |
| Sr | 3.38 | 3.22 | 3.00 | 2.84 | 2.71 | 0.72 |
| Al | 3.62 | 3.34 | 3.10 | 3.06 | 2.98 | 0.31 |
| Ti | 3.73 | 3.46 | 3.19 | 3.10 | 3.07 | 1.06 |

In Fig. 10, the $\Phi$ values of electropositive adsorbates, Li, Na, K, Rb, and Cs-doped graphene are presented. For single adatom-doped graphene (black solid line in Fig. 10), $\Phi$ decreases as we go down in the group. As we showed in Table II, doping by Cs results in the largest charge transfer to the graphene. Therefore, Cs-doped graphene has the lowest $\Phi$. By increasing the concentration, $\Phi$ of graphene is reduced further till it gets saturated. For Cs-, Rb-, K-, Na-, and Li-doped graphene, the saturation values of $\Phi$ are 2.05, 2.18, 3.24, 2.42 and 2.49 eV respectively. In the case of K-doping, the present results agree well with experimentally reported values (2.2 eV) [45].

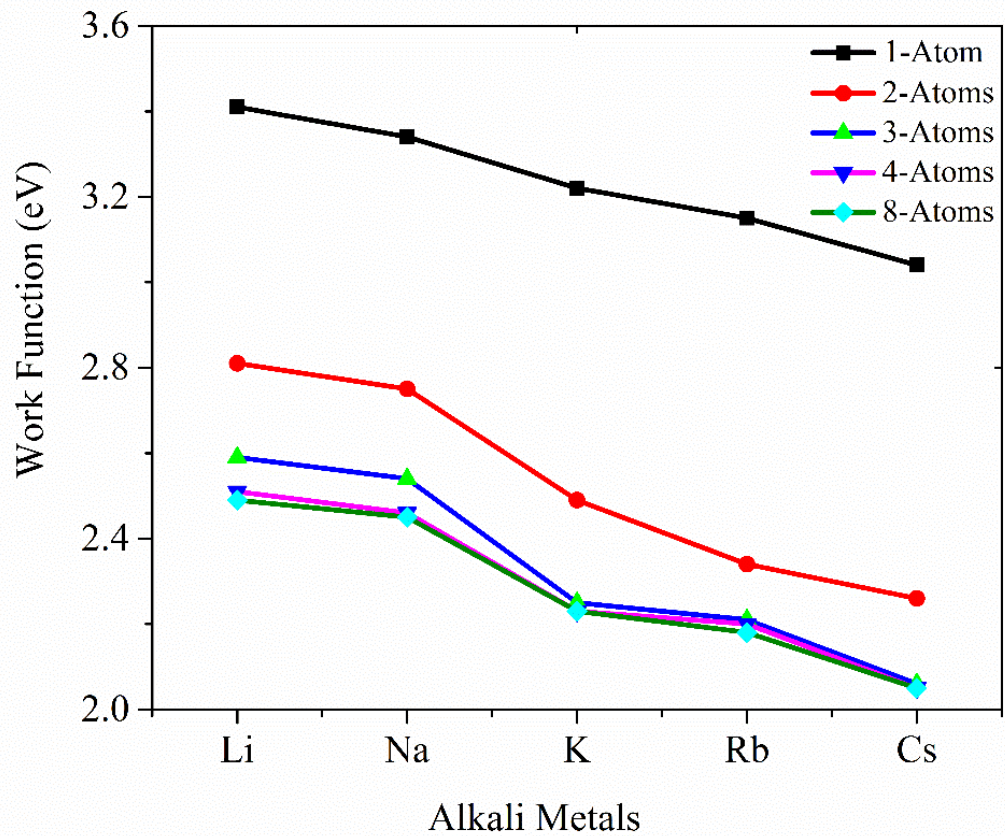

Figure-10    Work function ($\Phi$) of different concentrations of alkali metal-doped graphene.

## Conclusions

In graphene based electronic devices, the graphene work function is one of the most important properties that should be taken into considerations. In this work, the work function of graphene doped by different metallic dopants and different concentrations has been studied using DFT. In general, the work function of metallic doped graphene decreases with increasing the concentration of metal adatoms and saturated at certain values. For Cs-, Rb-, K-, Na-, and Li-doped graphene, the saturation values of $\Phi$ are 2.05, 2.18, 3.24, 2.42 and 2.49 eV respectively. The transfer of electron from metal dopants to the pristine graphene for all the studied systems shift $E_F$ above the Dirac point. As a result doped graphene becomes metallic. These Fermi energy shift is one of the main consequence of the significant reduction of $\Phi$.

## Acknowledgement

Funding for this research was provided by Qatar National Research Foundation (QNRF), grant no. NPRP 7-317-1-055. Computational resources are provided by research computing at Texas A&M University at Qatar.